\documentclass[10pt,conference]{IEEEtran}
\usepackage{hyperref}
\usepackage{times}
\usepackage{amsthm}
\usepackage{fixltx2e}
\usepackage{graphicx}
\usepackage{tabularx}
\usepackage{caption}
\usepackage{xcolor,colortbl}
\usepackage{mathtools}
\usepackage{array}

\IEEEoverridecommandlockouts

\begin{document}
\pagenumbering{gobble}
\title{\textbf{\Large Localization of real world regression Bugs\\[-1.5ex]using single execution}\\[0.2ex]\thanks{This research was supported in part by THE ISRAEL SCIENCE FOUNDATION (grant No. 476/11)}}

\author{\IEEEauthorblockN{~\\[-0.4ex]\large Dekel Cohen\\[0.3ex]\normalsize}
\IEEEauthorblockA{The Blavatnik School of Computer Science\\
	Tel-Aviv University\\Tel-Aviv, Israel\\
Email: {\tt dekelcoh@tau.ac.il}}
\and
\IEEEauthorblockN{~\\[-0.4ex]\large Amiram Yehudai\\[0.3ex]\normalsize}
\IEEEauthorblockA{The Blavatnik School of Computer Science\\
	Tel-Aviv University\\Tel-Aviv, Israel\\
Email: {\tt amiramy@tau.ac.il}} }

\maketitle
% \footnote{An example footnote.}

\begin{abstract}
%\boldmath
Regression bugs occur whenever software functionality that previously worked as desired stops working, or no longer works as expected. Code changes, such as bug fixes or new feature work, may result in a regression bug. Regression bugs are an annoying and painful phenomena
in the software development process, requiring a great deal of effort to localize, effectively hindering team progress.
In this paper we present Regression Detective, a method which assists the developer locating source code segments that caused a given regression bug. Unlike some of the existing tools, our approach doesn't require an automated test suite or executing past versions of the system. It is highly scalable to millions of loc systems. The developer, who has no prior knowledge of the code or the bug, reproduces the bug according to the steps described in the bug database. We evaluated our approach with bugs from leading open source projects (Eclipse, Tomcat, Ant). 
In over 90\% of the cases, the developer only has to examine 10-20 lines of code in order to locate the bug, regardless of the code base size. 
\end{abstract}


% \fi
\IEEEpeerreviewmaketitle

\section{Introduction}
Program evolution and repair are major activities of software maintenance, which consumes a significant fraction of the total cost of software production \cite{seacord2003modernizing}

During software development, new functionality is introduced, increasing the complexity of the system. 
Increased complexity often results in a reduced ability to estimate the impact of the code change. For example, mature and successful software, often supports several operating systems versions and configurations. When a developer makes a change, they are  not always aware of all the configurations impacted.

Further more, every code change now has to interact with more and more parts of the existing code, because it needs to reuse some of its functionality and integrate with  other parts. Often, the older code was not designed with many of the current requirements in mind, so parts may need to be rewritten, to allow reuse by the new code, or comply with non-functional requirements, such as security and performance. 

While code reuse is a desired property of a system, changes to infrastructure code can easily result in a break of one of its many clients (callers), especially when the interface (contract) is not well defined, and the clients hold different assumptions.

Sometimes, the maintainer is not the same person who wrote the code, which results in a loss of knowledge about system behaviours and flows \cite{yin2011fixes}. Being aware of the increased risk, the maintainers may employ minimal-change strategy, often referred to as \textit{patching}. While minimal-change (no refactoring) is low risk in the short term, it is a kind of greedy approach, often resulting in surprising or hard to understand code. Another outcome of no-refactoring strategy is code duplication, because the maintainer needs to reuse subset of the existing code, but doesn't want to change it. Code duplication in its turn is known to surprise future maintainers (sometimes even the same maintainer in the future). Typically an issue is fixed in some of the clones, but not in all of them.      

Some of the studies \cite{yin2011fixes} focus on the post-release bugs or security patches for stable slow-changing systems. Note that many of the regression bugs are introduced and fixed during the internal development process iterations, long before maintenance phase.

In general, every code change bears a risk of resulting in unintended behaviour, regressing functionality that previously worked as desired. To mitigate the risk, regression testing was adopted in an attempt to revalidate the old functionality inherited from the old version. Unfortunately, exhaustive regression testing is not always cost effective \cite{yoo2012regression,cibulski2011regression}, especially if it requires manual effort. Other mitigations, such as Code Reviews, Change classification \cite{kim2008classifying} and Change Impact Analysis using dynamic and static techniques \cite{ren2004chianti, orso2003leveraging} were also studied. Still, significant number of regression bugs are being fixed every day.  

Today, it is a time consuming task to locate the change that triggered the regression bug. From the author's experience, many times the developer is analyzing the bug as if it was a regular, non-regression bug. Others use the log history of the source control to try to identify the offending change, but may have to traverse large number of unrelated changes. Developers can also install previous versions of the system, to reduce the number of changes they must inspect. Installation and configuration, as well as reproduction of the bug, may be time consuming, unless they are automated, which is not usually the case for new bugs.

Our goal in this paper, is to explore a simple and scalable method that allows the developer to rank source code changes with respect to a specific regression bug scenario. The developer, who has no prior knowledge of the code or the bug, reproduces the bug according to the steps described in the bug database. As such, the tool requires only minimal training. As we understand that runtime overhead is an inhibiting factor in developer tools adoption, we kept it to a minimal percentage of the execution time.

In summary, we make the following contributions:
\begin{itemize}
\item Highly accurate localization for real world bugs, tested on several open source projects from different domains.
\item Simple to implement and reason about.
\item Demonstrates the temporal trace locality of the faulty change and the error, by using the novel \textit{Execution Order} ranking. 
\item Combines multiple weak filters and ranking methods in a novel way, to dramatically reduce the search effort required.

\item The requirements for tool operation are the same as for traditional debugging: Execute the bug scenario with the instrumented build. Automated test suites, special infrastructure or lengthy training processes are not required.

\item Usability: Complements the natural workflow of the debugging process, without requiring special knowledge.

\item Doesn't require executing past versions of the system, unlike previous work \cite{zhang2011localizing}
\item Minimal runtime overhead.
\item Locates "Not reproducible" bugs, that cannot be reproduced during debugging, due to date-time effects (ex: leap year, daylight saving time), thread schedule or any special configuration or input from the environment, that triggers the bug.   
\item Fix localization: When a certain version or branch of the software contains a fix that should be ported (merged) to another branch / version, our method can locate the fix change.  
\end{itemize}

\section{Our Approach}
We will now describe a method to rank source code changes that cause or expose a regression bug. 
Let's start with a more precise definition of the software regression bug that we try to localize.

\textit{Software Regression}: Software functionality that previously worked as desired and stops working, or no longer works as expected, given the exact same external environment.
Note that this definition requires a change in the system to trigger the regression bug, because all inputs to the system are held constant. 

As a starting point, the developer obtains a diff file between the last known good version of the software and the current, buggy version. The diff file is typically obtained from the revision control system, but can also be obtained by comparing 2 source file trees using a diff utility. The diff file format consists of a set of \textit{hunks}. A \textit{hunk} is a consecutive range of lines added, updated or deleted. After executing the code, each hunk contains one or more sub-ranges of code that actually executed. We rank the suspect changes at the hunk sub-ranges granularity, which usually means few consecutive lines. We will refer to it from now on, as a \textit{change} or \textit{code change}.
Note that this method assumes that one or more of the changes in the diff is the root cause, or the exposing change for the bug. Note also that while the number of changes from the good version vary, it could reach hundreds or thousands of changes. The tester or customer who opened the bug, specifies the previous release (say a year ago) as the good version and does not know the exact revision that introduced the bug, which increases the size of the diff file considerably.

We use a combination of dynamic analysis methods and heuristics to rank the source code changes.
\subsection{Executed Changes}
The first filter, is executed change - a hunk that was not, at least partially executed in the bug reproducer scenario is not displayed to the user in the search results.

In a 1 year diff of an active large system, this filter, by itself, can typically reduce the number of changes from thousands to hundreds. 

\subsection{Execution Order} 
Execution Order is a light weight heuristic, assuming temporal locality between faults and observed errors. Preliminary  investigation indicated, that many of the regression observed errors are in a short trace-distance from their root-cause faulty change. Let's define the executed trace T of a program P as the ordered sequence of basic blocks {B\textsubscript{j}}. \[<B_{i_1},B_{i_2},B_{i_3},...>\] \textit{trace-distance} is defined between 2 elements of the trace , as the difference of positions in the sequence between the two elements. In the example above: \[tracedist(B_{i_1},B_{i_3}) = 2\]
A hunk H may span several basic blocks. Our method instruments all basic blocks of a hunk to log its execution. The result is an ordered sequence of executed hunks (with repetitions) \[<H_{i_1},H_{i_2},H_{i_3},...>\] 
The developer is instructed to save execution order state as soon as possible after error is observed. The tool offers several ways to accomplish that. For example, if the bug results in an exception, it is easy to stop in the debugger on exception throw and then save the execution order. If it is a hang or any incorrect result in UI, the developer can click the 'dump' button right after (see \ref{workflow}). Often it is easy to place a breakpoint to suspend execution sometimes after the error occurred and then save (dump). Although it is important to dump the state as soon as possible, we didn't observe high sensitivity, especially if combined with the \textit{Differential Basic Block Hit} ranking methods (see next section). We then employ a simple ranking (that can be combined with other ranking scores as well): The last hunk logged before the dump is ranked first (1), the second hunk further away is second and so on, according to the trace-distance from the dump point. 

The Execution Order method allows us to effectively rank dozens, or even hundreds of hunks, with the faulty hunk often near the top list. We provide here the intuition behind this method: Today, one of the most helpful supportive information, developers use to debug is a stack trace \cite{schroter2010stack}. When available, the developer browses the stack frames preceding the error location in order to search for the fault. Execution Order covers this scenario and more: It also locates a change, even if it is not in the active stack trace, but has a relatively short trace-distance from the error. Note that while stack trace can be extracted mainly for bugs of type crash and hang, Execution Order is almost always available, because one can always dump the state after the error occurred.
  
\subsection{Differential Basic Block Hit}
There is a problem with \textit{Execution Order}: If other, non faulty changes, are executed after the faulty change and before the state is saved, the top ranking changes will not be the faulty ones we are after. During preliminary research, the problem presented itself mainly in 2 typical cases: a) Changes in a background task, executing concurrently with the bug scenario and b) Changes in low level code that executes often, such as UI message loop code. If a breakpoint is set right after the error, the interference is insignificant, but in those cases the developer is using the UI button to dump the state, while changes continue to execute for another second or so, an additional ranking or noise filtering is required, in order to achieve top ranks. 

As a solution, we further isolate the changes relevant to the bug scenario by using \textit{Differential Basic Block Hit}.  
\begin{enumerate}
\item The code is instrumented to collect basic blocks hit coverage information.
\item The developer first executes the code in a scenario that doesn't reproduce the problem and then saves the state of the covered basic blocks. This is usually done after the system has initialized, before the developer starts reproducing the bug.
\item Then the bug scenario is executed and then state is saved again (cumulative blocks coverage + \textit{Execution Order} from previous section). 
\item Difference operator is used to find all basic blocks that were executed in the bug scenario, but not in the first coverage dump (non bug scenario).
\item We rank first all changes that have in their vicinity (in the same method, max 10 lines distance) a basic block from the coverage diff, and all other changes after. Note that within this rank, we combine a secondary \textit{Execution Order} rank, described above.
\end{enumerate}
 
Note that this is not the same thing as ranking first the changes that were executed in the bug scenario but not in the non-bug scenario. Initial investigation revealed that even if the change was executed in both scenarios, but is in the same method, at a certain distance from a basic block in the diff, it should be ranked first, because it may be in the backward slice of a conditional that has different value in the 2 scenarios, therefore affecting the buggy control-flow.

Sensitivity is also an issue here: What if the developer recorded the wrong non-bug scenario ?
The main risk, is that the faulty change will enter also the non-bug coverage (instead of only to the bug scenario), lowering it in the rank. In practice, we didn't encounter difficulties in this case, because we only use the non-bug scenario for "background noise filtering" purposes, so it doesn't have to be very similar to the bug scenario, lowering the risk. Typically system activity is recorded to include the background periodic tasks and low level common code in the non-bug coverage. Often the non-bug recording doesn't activate the product feature which the bug occurred in. For example: For a bugs in a search feature of a product, launch the instrumented system and perform few operations, without activating search. This is our non-bug scenario that will isolate only those changes related only to search functionality.

This method alone can reduce in some cases the number of ranked-first changes to several dozens. In combination with the \textit{Execution Order}, we get the final results.
Note that \textit{Differential Basic Block Hit} can be used as a filter or as a ranking method, where the changes that occur in methods with difference are ranked before those that occur in methods with no difference.

\subsection{Semantic Textual Affinity}
Textual search is used on a daily basis by developers (including the authors), usually via the Editor Find functionality or command line tools, such as grep. There are several well known problems when using grep like tools:
\begin{enumerate}
\item Lack of ranking: If the search term appears in method C multiple times, the search result will not be presented before a hit in method B, where it appears only once. As another example, term hit in method name is not ranked higher compared to a hit in method body. 
\item Search terms (or regular expression patterns) must generate exact hit, so searching for the term \textit{login} does not generate a hit if the code only contains related terms such as \textit{authentication} or \textit{password}.
\item Program structure implying distance between terms is not utilized. For example, if the search term occurs in a callee of method containing the faulty change, which resides in a different file, no hit is generated by grep.
\end{enumerate}
We investigated and implemented a novel textual affinity search engine as an additional ranking method, combined with the dynamic analysis methods described above. The basis for the Affinity search was laid by \cite{nir2008locating} and \cite{cibulski2011regression} and relies on several IR and program structure ranking heuristics. The developer uses search terms, which may be identical or related to the actual terms appearing in or near the faulty change.

\begin{enumerate}
\item Identifier names in the source code and comments are split into parts based on known coding standards, such as CamelCase and underline whereas the original form of each identifier is preserved as well.

\item Java keywords are treated as stop words, in addition to standard English stop words list.
\item For matching search terms that do not appear in the source code, we have used 2 different methods: For extract terms appearing in WordNet lexicon \cite{miller1995wordnet}, we use an an adapted version of the CodePsychologist’s affinity score algorithm, which determines the score as a function of the inverse distance in WordNet taxonomy tree. The tree represents relations such as synonymy and hypernymy. See more details in \cite{nir2008locating}. 
\item For other terms, that are not covered by WordNet, we use Snowball stemmer to determine equivalence. 
\item WordNet does not specialize in software engineering terms, so the relatedness score between two related terms, such as \textit{screen} and \textit{resolution} is surprisingly low. To overcome this limitation, we also experimented with collocation techniques, where terms that co-occur together often in software engineering corpus are considered as related. Our choice for corpus was Stack Overflow \cite{stackoverflow}, where we used titles and tags of a subset of the posts. The initial experiment results are not conclusive.
\item Term Weighting is an important part of current IR technology. We created a weighting scheme as follows:
  \begin{enumerate}
  	\item TF-IDF: TF-IDF assign higher weight to term hits that appear frequently in few files and do not appear frequently in other source files. We have used a robust variant of TF-IDF adapted to counter problems of frequency imbalance.
  	\item Since the targets of the search are changes, term hit is assigned higher weight if inside an executed change, compared to changes further away (in lines of code) from the executed changes. 
  	\item Term hit in the containing method name, its parameters or the method comments are also assigned higher weight, as well as the containing class name.
  	
\end{enumerate}

\item We have also experimented with static Callgraph \cite{wala}, to allow for hits in methods called from the method containing the change. these are assigned lower weight as their Callgraph distance is longer. We also limit the search to 3 levels of caller-callee.
\item Finally, the search rank is combined with the other methods to form the final rank.
\end{enumerate}

We ended up omitting \textit{Textual Affinity} from the experiments because of the following reasons:
\begin{enumerate}
\item In our experiments, the results were satisfactory without using \textit{Textual Affinity}.
\item Validation complexity: It is harder to objectively choose the search term. In fact, we discovered it is only applicable to subset of the bugs, which include relevant terms in their Titles or Description. The same or related terms should appear in the vicinity of the faulty change, which is not always the case. Still it may be useful, as we plan to investigate in the future.

\item We still need to enhance and improve the functionality of this engine.
\end{enumerate}

\subsection{Developer workflow} \label{workflow}
To get a better grasp of the ranking methods described and also review the user interface of the Regression Detective tool, we will go through an example, step by step.
The tool is implemented as Eclipse plugin and currently supports Java.
\begin{enumerate}
\item The developer starts with steps to execute the bug scenario. Note that formal bug report is not required. In our example the bug is from the Apache Ant project \href{https://issues.apache.org/bugzilla/show_bug.cgi?id=52923}{Bug 52923: JUnit4 test should not run as JUnit3 if annotated RunWith(JUnit4)}
\\[0.25cm]
\setlength\fboxsep{10pt}
\setlength\fboxrule{1pt}
\fbox{\includegraphics[width=6cm]{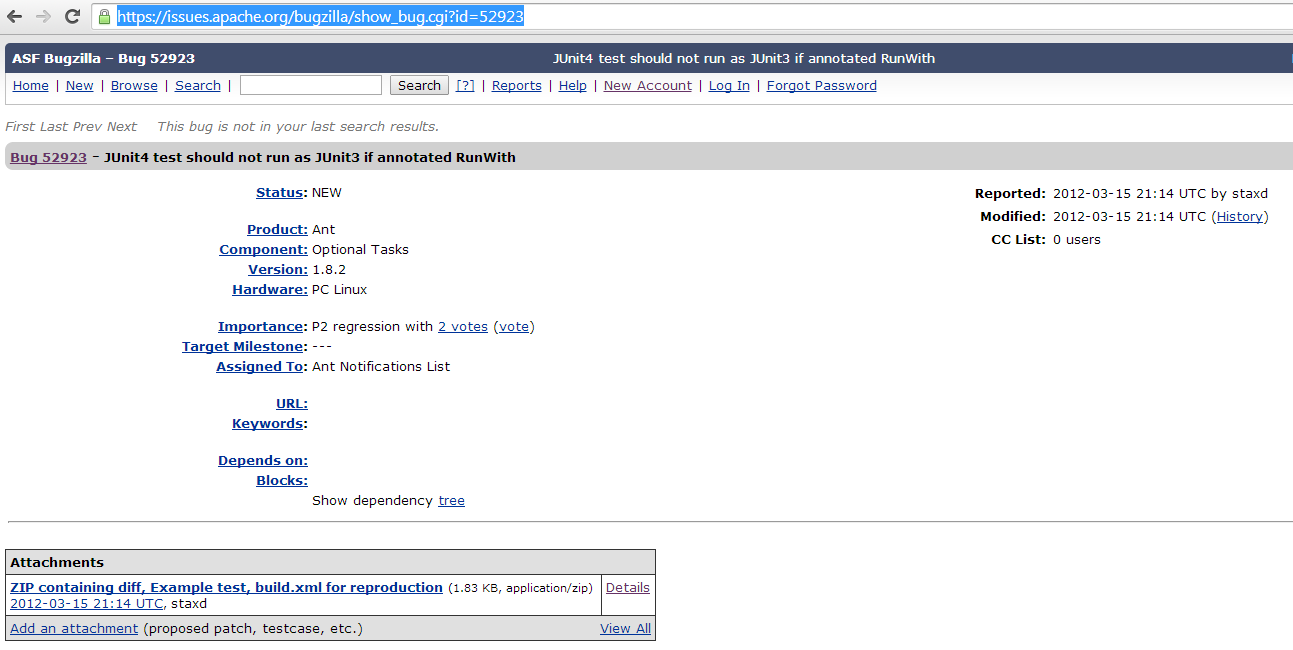}}
\\[0.25cm]
\item Next, the last known good version is determined. A diff file is created that includes all changes from the good version to current, buggy version. The developer typically uses revision control system, such as Subversion or Git (not required). The diff file is imported into the IDE using a wizard dialog. Our bug is a regression from ant 1.8.1 reported and fixed at ant 1.8.2
\\[0.25cm]
\fbox{\includegraphics[width=6cm]{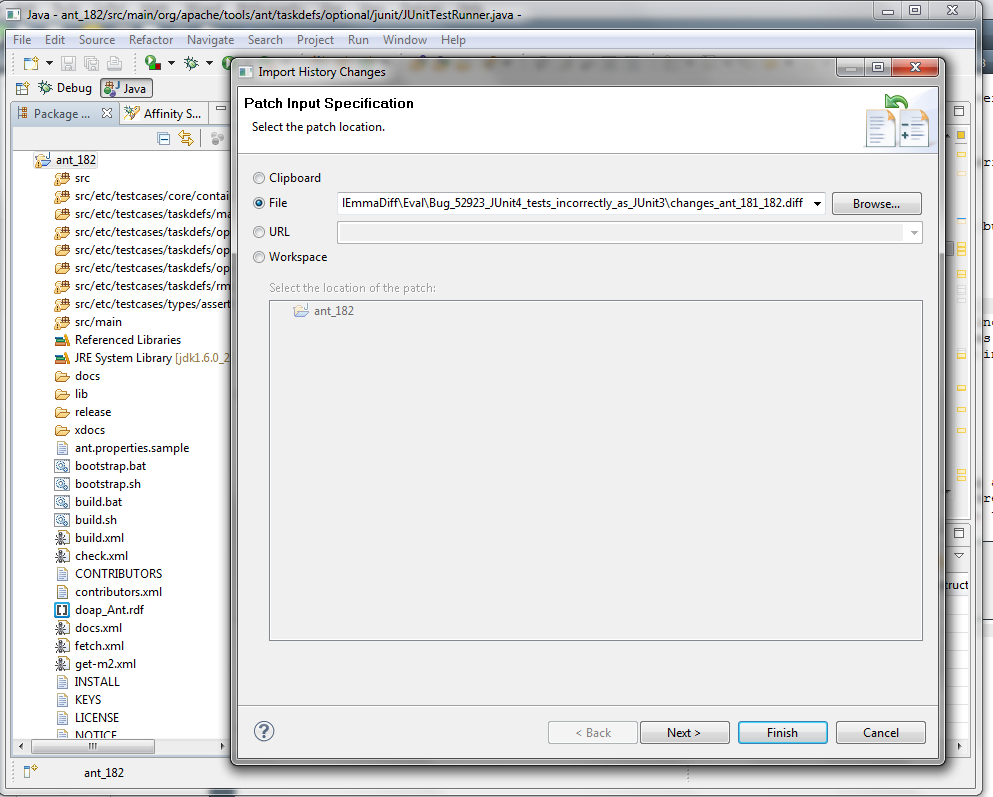}}
\\[0.25cm]
\item The developer can optionally instrument the changes to activate \textit{Execution Order}.  When the program is launched, it is anyway instrumented for basic blocks coverage (used for \textit{Executed Changes} and \textit{Differential Basic Block Hit} methods, described above. 
\\[0.25cm]
\fbox{\includegraphics[width=6cm]{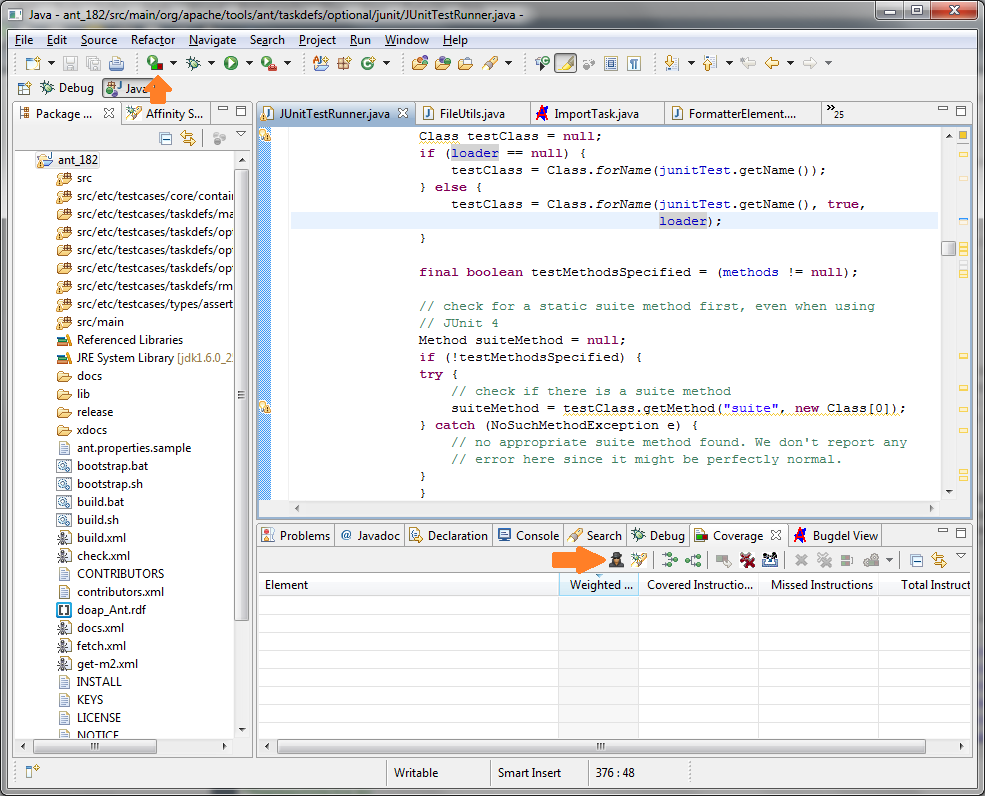}}
\\[0.25cm]
\item Now the reproducer test (attached to the bug) is executed and the developer clicks the dump button (in the same toolbar shown in the image above) to save the state of the Execution Order and the covered basic blocks. Note that ant is a command line process, hosting JUnit that in turn loads and executes our user-code test (using the wrong JUnit version). Since the whole runtime takes a second and it may be hard to place a breakpoint in this case inside the test, the developer can easily add a sleep() statement to to their own test code, allow enough time to click the Dump button.
\item After the execution data was collected, our plugin displays search results in a view, which resembles popular search engines UI, except that in this bug, we don't need textual search, so we leave the search term fields empty, simply clicking the search button. 
\\[0.25cm]
\fbox{\includegraphics[width=6cm]{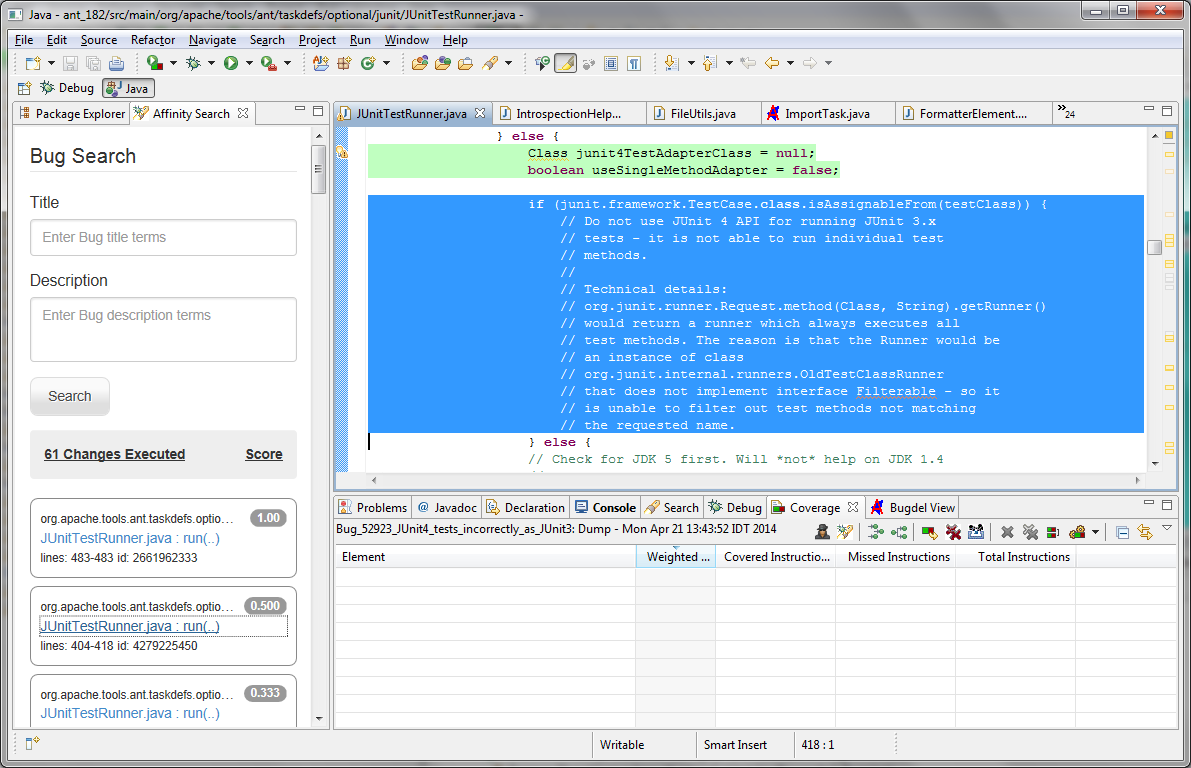}}
\\[0.25cm]

\item It can be seen in the search view, that the second result is the faulty change. When clicked, it is selected in Eclipse Java code editor.
\end{enumerate}
Note that in this specific bug, the \textit{Execution Order} is sufficient to localize the bug, and further ranking with \textit{Differential Basic Block Hit} or \textit{Textual Affinity} was not required, since we already have a ranking of 2. The developer can incrementally choose to apply further ranking methods, as needed.

The tool is integrated with the workflow of the developer and adds only few clicks to the traditional debugger workflow. after clicking a search result to inspect it in the Java Editor of Eclipse, she can either fix it or place breakpoints to continue the investigation. She can also examine the executed code (see green highlighted lines in the above screenshot) vs. the non-executed (in red) which helps to focus on the relevant area.

\subsection{Not reproducible bugs}
All the change ranking methods described above do not rely in any way on actually reproducing the error in the developer environment. It is sufficient that the faulty change is executed close to the dump state point. If \textit{Differential Basic Block Hit} is used, it may be sufficient that the the faulty change is near instructions that were executed in the bug scenario but not in the non-bug scenario. Note that by actually reproducing the bug, the developer gains high confidence that they are executing the right scenario, but this is not the tool or the method requirement. Even if the order of the changes executed is slightly different (as long as the above sufficient conditions are met), the method is still effective. 

One of the authors witnessed such a real world customer  regression bug, that couldn't be reproduced in the R\&D labs. After investing several days to no end, a senior architect (who was also the original developer of the code) was flown to the customer site to investigate the bug on a specific machine, where it can be reproduced. The root cause was a race condition, as a result of a code change. The faulty change was always executed close to the error point. It was determined, by manual analysis  that the bug could be easily detected by our prototype tool, except it doesn't support the  runtime and language used. Today, Regression Detective only supports Java, but other runtime and language support can be added. 

Note that we do not claim that the above methods are optimal for localizing race conditions as there is a large body of work specializing in this subject. There are other common reasons for not reproducing the bug, while still meeting the above sufficient conditions, such as different external environment : date time settings, locale settings, interaction with other software - installed only on the reproducing machine and so on.

Unfortunately we didn't find not reproducible bugs regression in the open source repositories used for our evaluation, but our methods are easy to reason about, in order to determine such a property, as discussed above.

\subsection{Fix localization}
Software development processes often result in a creation of branches, which represent different versions of the system \cite{appleton1998streamed}. It is sometimes the case, that a bug was fixed or a feature developed in one of the branches, without merging to all other branches immediately. When the need arises to port a specific bug fix or functionality to the other branches, the developer has to first localize the relevant changes in the source branch. It is often the case that the merge includes changes not documented in known bugs or specifications. Since our tool doesn't assume the changes are necessarily related to bugs, it can be used to find any type of changes (fix, new functionality). The developer chooses a scenario that executes the changes and use our ranking methods to locate the change.
Another common scenario is fix reuse. The authors have recently encountered a bug, which is not a regression. Its scenario and control flow is very similar to a another, already fixed bug. The developer assigned with the new bug was not aware of the previously fixed bug with the similar scenario, in the same area in the code. With our tool, she can localize the existing fix code, in the new bug scenario, which can be reused to handle also the new bug.
In our example, the already fixed bug was rewriting a certain absolute url. The new bug was similar, but with a relative url. Why is it considered a common case ? Because often bugs are reported and fixed for a certain scenario, while other related defective scenarios are not fully considered. Sometimes this partial fix results in a regression. In other cases, the fix only handles the exact reported scenario and should be extended to handle also the related similar defective scenarios.

\section{Implementation}
We implemented Regression Detective as Eclipse \cite{eclipse} plugin for the Java programming language. Java is a popular programming language with several leading Open Source projects we could use for evaluation. Eclipse is a popular IDE with plugin extensibility architecture, used in much of the academic related work. We used a modified version of Emma \cite{emma} and EclEmma \cite{eclemma} Eclipse plugin as the basis for our coverage collector and UI, and added BugDel \cite{usui2005bugdel} with JavaAssist \cite{chiba1998javassist} for hunk instrumentation. The textual affinity search engine was based on \cite{cibulski2011regression} code with numerous modifications. Eclipse JDT model and its compare plugin allowed us to reduce the amount of code we had to write. 

Regression Detective uses 2 types of instrumentation: The first is Emma coverage for basic blocks, which sets a boolean flag to true when the block is executed. The second one is hunk instrumentation, which logs in a memory circular buffer the hunk id, whenever it is executed.

The Search results view was created in Html5 JavaScript technologies using AngularJS  \cite{angularjs} and is an interesting example of integrating Web technologies in a Desktop application - Eclipse. Such an approach will enable our team to port the project to different IDEs in the future.

\section{Experiment}

\begin{table}
\begin{tabular}{|m{2cm}|m{4cm}|c|}
\hline
$Name$&$Description$&$LOC$\\
\hline
Apache Tomcat&Java application server, widely adopted for large-scale, mission-critical web applications&~1M\\
\hline
Eclipse JDT Core&The Java tooling infrastructure of the Eclipse Java IDE&~1.5M\\ 
\hline
Apache Ant&Build system command line tool&200K\\
\hline
\end{tabular}
\caption{Open source projects used for evaluation}
\label{tab:evalprojs}
\end{table}

One of our main objectives was to demonstrate localization of real world regression bugs.
Therefore, we selected several leading open source projects representing different application domains, such as Web application server, IDE compiler, program manipulation library and command line tools - see table \ref{tab:evalprojs}. Note that the number of lines of code refers to the versions actually used in the evaluation, as some of these projects have grown a lot larger over the years. 

Next, we randomly selected regression bugs from the projects bug databases. Many of the regression bugs are not clearly marked as such, which affected the number of bugs that could participate in the evaluation. 

For each bug, we setup a development environment of the source code and build system in Eclipse for the version when the bug was reported and fixed. We encountered technical difficulties in some of the older bugs (for example a bug in Apache Tomcat 5.0.1) which forced us to omit several bugs, whose build dependencies were hard to acquire or the build instructions were missing.

Bug reports often include the regression and fix location in the description or the comments. Information extracted before the experiment consist of the title and part of the description that doesn't include the fix location. It is important to simulate the common case, where the developer doesn't already know the location of the bug.

It is interesting to read some of the discussions in the comments which indicate that it sometimes require days of collaboration between several developers, users and testers to localize such regressions. 

We reproduce the bug, saving dumps of the \textit{Execution Order} and often also coverage information for \textit{Differential Basic Block Hit}. The precise point for save dumps was selected per bug and is detailed in table \ref{tab:evalbugs}.  

For ground truth we used the fix associated with the bug. Many of the bugs have a patch with the fix attached, or a fix revision in the revision control system associated with the bug. We had to disqualify bugs which didn't have a clear fix location.

\section{Results}

\definecolor{Gray}{gray}{0.85}
\newcolumntype{a}{>{\columncolor{Gray}}m{1cm}}

\begin{table*}[!h]

\begin{tabularx}{\textwidth}{|m{1cm}|m{1cm}|X|m{0.5cm}|m{0.5cm}|X|}
\arrayrulecolor{black}\hline
\rowcolor{Gray}
\hline
  \textbf{Bug} & \textbf{From}  & \textbf{Description} & \multicolumn{2}{ c }{\textbf{Rank}} & \textbf{Dump Point}\\

\rowcolor{Gray}

& & &  \textbf{EO} & \textbf{EO+D} &\\
\hline

\href{https://issues.apache.org/bugzilla/show_bug.cgi?id=52923}{52923}&\small Ant&\small JUnit4 test should not run as JUnit3 if annotated RunWith(JUnit4)&2&-&\small Added delay inside the reproducer test case (Example.java) to sleep for few seconds to allow clicking dump button after the (wrong) test method was called\\

\hline

\href{https://issues.apache.org/bugzilla/show_bug.cgi?id=50007}{50007}&\small Ant&\small Taskdef classpath breaks with directory containing an exclamation mark (!)&4&-&\small When exception is thrown\\

\hline

\href{https://issues.apache.org/bugzilla/show_bug.cgi?id=50953}{50953}&\small Ant&\small Tasks defined with an antlib within a jar in succeeds with Ant 1.7.1 and fails with 1.8.2&1&-&\small When exception is thrown\\

\hline

\href{https://issues.apache.org/bugzilla/show_bug.cgi?id=51387}{51387}&\small Ant&\small Performance regression in ant task to launch external processes&2&-&\small Used VisualVM Profiler to automatically locate the hotspots, then paused in debugger\\

\hline

\href{https://issues.apache.org/bugzilla/show_bug.cgi?id=55227}{55227}&\small Ant&\small JUnit 3 tests $\xrightarrow{}$ run with JUnit 4 $\xrightarrow{}$ put @Ignore annotation $\xrightarrow{}$ ant incorrectly executes the @igonored test&2&-&\small Added delay inside the reproducer test case (Example.java) to sleep for few seconds to allow clicking dump button after the (wrong) test method was called\\

\hline

\href{https://bugs.eclipse.org/bugs/show_bug.cgi?id=252887}{252887}&\small Eclipse&\small Pressing F1 causes editor properties to disappear&10&6&\small Breakpoint in main keyboard handler after pressed F1\\

\hline

\href{https://bugs.eclipse.org/bugs/show_bug.cgi?id=378390}{378390}&\small Eclipse&\small Java Search regression for references of a method&1&1&\small Recorded Eclipse Java Editor with another file + Open Search dialog, but didn't execute search $\xrightarrow{}$ Dump $\xrightarrow{}$ reproduce 0 results Java Search $\xrightarrow{}$ Dump again\\

\hline

\href{https://issues.apache.org/bugzilla/show_bug.cgi?id=18201}{18201}&\small Tomcat&\small getReader() does not throw UnsupportedEncodingException when bogus charset is used&1&1&\small Breakpoint after getReader (which is called directly from user servlet test reproducer code)\\

\hline

\href{https://issues.apache.org/bugzilla/show_bug.cgi?id=44405}{44405}&\small Tomcat&\small Regression with tomcat 6.0.16, NullPointerException on getServletContext().getResourceAsStream()&13&1&\small Recorded several requests $\xrightarrow{}$ Dump $\xrightarrow{}$ reproduce NullPointerException $\xrightarrow{}$ Dump\\

\hline

\end{tabularx}
\caption{Evaluation results}\label{tab:evalbugs}
 \caption*{
 	\begin{tabular}{l l}      
      EO: & Execution Order\\
      EO+D: & Execution Order + Differential Basic Block Hit 
    \end{tabular}
 }
\end{table*}

As can be seen in the Evaluation results table, the ranking achieved using a combination of \textit{Execution Order} and \textit{Differential Basic Block Hit} is 1-6.

During the experiments, we observed that if one or more of the faulty hunks are ranked in the top 10 search result entries, the developer can quickly browse and locate the offending change in a matter of minutes.

Note that for the first few bugs (in Ant project) only \textit{Execution Order} was used, mainly because the developer didn't need \textit{Differential Basic Block Hit} ,as the ranking is already high.
\section{Runtime overhead}
When developers use debugging tools in a lab environment, they can usually tolerate a minor slowdown, such that their workflow is not disrupted.
We have measured a 1-10 percent slowdowns, depending on the test:
\begin{enumerate}
\item Changes instrumentation overhead depends on number of changes and their location, but generally has a very low impact, almost not measurable. It inserts a function call that does very little work and writes a 64 bits long integer to a circular memory buffer. This is less than most function calls in the system and is usually quite rare in the execution trace (even compared to standard logging). 
\item We believe we can apply the changes instrumentation even in Production Servers, where the changes logging are taken to further analysis in the lab, where the developer can execute the same scenario with more instrumentation and discover the faulty change, even if the bug doesn't reproduce in the lab.
\item Much of the slowdown is in the Basic Blocks (Emma) instrumentation, as with any method that uses granular coverage for testing or fault localization. Our prototype is not at all optimized: We believe we could obtain similar results using only function entry logging, in addition to changes logging, which would reduce the overhead considerably.
\item Using our prototype with many real world bugs, our subjective impression is, that the developer can still debug at normal speed, with somewhat longer startup time for large projects (such as Eclipse). The slowdown is hardly noticeable with shorter scenarios, such as ant command line reproducer samples taken from the Ant bug database. 
\end{enumerate}

\section{Threats to Validity}
Our study, like any of this nature, has limitations which may impact the validity of our findings. 
We now proceed to identify some of these limitations, their impact, and how we attempted to address them.
The primary threat to the external validity is the generalizability of our results acquired from the 9 evaluated bugs in 3 Java based systems. Our Temporal trace distance assumption may not hold for example, for initialization bugs. In these bugs the faulty change is executed at system start up, but the error occurs much later.
In response, we intend to introduce slicing and dataflow based methods, to detect a data dependency between the error location and the faulty change, provided the bug scenario exhibits such location (Ex: Exception throw location). Our Textual Affinity search may also help with such cases. Note that if the faulty change occurs in a specific scenario, long before the error and doesn't occur in common scenarios, such as initialization, we can still localize it using the \textit{Differential Basic Block Hit} method.

\section{Related Work}
Several studies compare behaviors of faulty and correct program versions on the same inputs, using various spectra, such as path and value, to localize regressions \cite{xie2005checking,zhang2011localizing}
 
In our work, we used block hit spectra, only as a weak rank/filter, to isolate the area in code relevant to the regression. The actual offending lines are usually detected by ranking the program changes using both execution order, which is not used by the above.  The above methods also require instrumentation and execution of the faulty version (as in our work), plus the previous correct version, which our work doesn't require. While in some cases it is easy enough to obtain the previous versions, instrument and execute them, there are often changing configurations and external dependencies, such as remote services and database schemes that makes it difficult to setup older code, several months old, for debug. In addition, some of our evaluation subjects are much larger (Tomcat,Eclipse JDTCore), in terms of the amount of changes to rank, compared to most of the subject projects in the above work.

The CodePsychologist,
which was previously developed in our group \cite{nir2008locating}, is a tool which assists the programmer to locate source code segments that cause a given regression bug. The CodePsychologist uses affinity ranking to estimate how close a segment of code is to a given test case. Affinity between groups of words is calculated based on the semantic similarity between pairs of words from each group, measured as the inverse path length in a WordNet taxonomy.
While in our current work, we did also experiment with text based semantic similarity, we discovered that a) The bug descriptions do not always contain the search terms needed to locate the offending code change. Test case descriptions were used for search in \cite{nir2008locating} and b) That dynamic analysis by itself is accurate enough in most cases. Semantic search was successfully used in feature and bug localization in \cite{nir2008locating} and \cite{poshyvanyk2007feature} experimenting with range of IR techniques, such as LSI.

Darwin \cite{qi2012darwin} uses symbolic execution to automatically synthesize a new input that (a) is very similar to the failing input, and (b) does not fail. It then identifies code fragments where the control flow diverge. Darwin uses powerful techniques that are accurate and can even localize non-regression bugs. Unlike our method, these techniques incur significant runtime overhead, are a lot harder to implement, require reproduction of the bug and a reference system, where the bug doesn't reproduce.

RPRISM \cite{jagannathantech} localize regression faults using trace difference using two versions of the program (good,buggy) and three comparisons: correct scenario in the two versions (low rank) and regression scenario, which is correct in the good version and generates error in the buggy version (high rank) and diffs occurring between the correct and buggy scenario, only on the new version - this is similar to the diff we use in RegressionDetective. It aligns and compares ordered traces per thread, method and object-instance. Note that it doesn't exploit temporal trace distance, as in our work.

Software reconnaissance \cite{wilde1995software} is an early work that popularized using coverage differences to location features in code using 2 or more scenarios, some executing the feature code and some not. In our work, we use a similar method to localize the feature (or system module) containing the regression fault, filtering out irrelevant changes (noise).

Delta Debugging \cite{zeller1999yesterday} is divide-and-conquer greedy algorithm, systematically searching for failure-inducing changes by patching a subset of changes and observing execution results. It is capable of locating regression bugs, even in non-source code changes (such as configuration change) and is fully automated. On the other hand, it requires automated test, which may be hard to write for UI applications and an automated way to build and run the system in every past revision, which is non-trivial in many projects and may take hours (as explained above). As a greedy algorithm, it may get stuck in a local minima, outputting sub-optimal results. Unlike our method, it can only handle reproducible bugs. Recent study \cite{jss} evaluated delta debugging using real world regression bugs of unix command lines tools and concluded that two-thirds of the reported changes were related to the bug. 

RADAR \cite{pastore2012dynamic} localize regression faults using by constructing models of the good version and applying them on the trace of the new (bad) version, to detect model violations. It instruments only statements that occur unaltered in both the good and bad version of a modified function, plus its callers and callees. The data collected consist of Daikon \cite {ernst2007daikon} invariants and FSA to capture intra function control flow. Unlike our tool, RADAR relies on both 2 versions and a test suite of passing and failing tests. In the new version, only failing test cases are instrumented. Detected anomalities are reported in the order they are observed to allow easy understand of sequence of events. This is also very different from our approach, which reports the suspected changes closed to the observed error first.

FIFL \cite{zhang2013injecting} enhances FaultTracer \cite{zhang2011localizing} using a novel form of injected program edits, borrowed from Mutation Testing techniques. It assumes test suite with failing and passing tests, old version and new, buggy version of the code. It injects artificial mutants into the old version, rerun the test-suite. If there is a correlation between the tests results of the old code plus mutants and the new code (with bugs), it maps the mutants to real changes to boost their suspiciousness rank.

Fault Localization using dynamic slicing and change impact analysis \cite{alvesfault} uses a combination of backward and forward dynamic slicing from an incorrect test output value and from the change to filter out statements which do not affect the final result, or couldn't be affected by the change. It experiments with several accurate but high runtime overhead methods. It also assumes the presence of an incorrect output value to slice from.

\section{Conclusions and Future work}
We introduced a method to localize regression bugs that is not automatic, but requires very little from the developer, in comparison with the detailed know-how required to localize a bug with a traditional debugger. We assume that given a bug report, the developer should be able to reproduce the bug, or at least execute the bug scenario without reproducing it, exercising the faulty change. The developer also has to click the "dump" button twice, before and right after the error is observed. In our evaluation, it only took several minutes to obtain a ranked search result for a given bug. Our main result, is that combining temporal locality between faults and observed errors, with program execution differences between the regression scenario and a non-bug scenario can help the developer rank the faulty changes. We also created a Textual Affinity search engine as another ranking methods, but ended up omitting it from the experiments.

We view Delta Debugging as potentially complementary to our approach: The developer can benefit from experimenting with our top rank suspected changes, and perform some automatic or semi-automatic patching to further isolate the faulty change. The iregression project \cite{iregression_ase12}  filters out non-executing changes as preprocessing step for its Delta Debugging algorithm variant.

Our current method only handles changes to executable code. For example, faulty change to a configuration file is not taken into account. We investigate scalable dynamic data-flow analysis methods to alleviate this problem in the future.

Refactoring changes, such as local variable rename, rarely result in a regression bug. Work on identifying refactoring changes \cite{weissgerber2006identifying} can be used to improve our ranking heuristics.
 



%\bibliographystyle{IEEEtran}
% As suggested below, edit bibtemplate_samples.bib to reflect
% your bibliography. See bibtemplate.text for referencing.
%

\bibliographystyle{IEEEtran}
\bibliography{dekelbibfile}

\begin{thebibliography}{10}
\providecommand{\url}[1]{#1}
\csname url@samestyle\endcsname
\providecommand{\newblock}{\relax}
\providecommand{\bibinfo}[2]{#2}
\providecommand{\BIBentrySTDinterwordspacing}{\spaceskip=0pt\relax}
\providecommand{\BIBentryALTinterwordstretchfactor}{4}
\providecommand{\BIBentryALTinterwordspacing}{\spaceskip=\fontdimen2\font plus
\BIBentryALTinterwordstretchfactor\fontdimen3\font minus
  \fontdimen4\font\relax}
\providecommand{\BIBforeignlanguage}[2]{{%
\expandafter\ifx\csname l@#1\endcsname\relax
\typeout{** WARNING: IEEEtran.bst: No hyphenation pattern has been}%
\typeout{** loaded for the language `#1'. Using the pattern for}%
\typeout{** the default language instead.}%
\else
\language=\csname l@#1\endcsname
\fi
#2}}
\providecommand{\BIBdecl}{\relax}
\BIBdecl

\bibitem{seacord2003modernizing}
R.~C. Seacord, D.~Plakosh, and G.~A. Lewis, Modernizing legacy systems:
  software technologies, engineering processes, and business practices.\hskip
  1em plus 0.5em minus 0.4em\relax Addison-Wesley Professional, 2003.

\bibitem{yin2011fixes}
Z.~Yin, D.~Yuan, Y.~Zhou, S.~Pasupathy, and L.~Bairavasundaram, ``How do fixes
  become bugs?'' in Proceedings of the 19th ACM SIGSOFT symposium and the 13th
  European conference on Foundations of software engineering.\hskip 1em plus
  0.5em minus 0.4em\relax ACM, 2011, pp. 26--36.

\bibitem{yoo2012regression}
S.~Yoo and M.~Harman, ``Regression testing minimization, selection and
  prioritization: a survey,'' Software Testing, Verification and Reliability,
  vol.~22, no.~2, 2012, pp. 67--120.

\bibitem{cibulski2011regression}
H.~Cibulski and A.~Yehudai, ``Regression test selection techniques for
  test-driven development,'' in Software Testing, Verification and Validation
  Workshops (ICSTW), 2011 IEEE Fourth International Conference on.\hskip 1em
  plus 0.5em minus 0.4em\relax IEEE, 2011, pp. 115--124.

\bibitem{kim2008classifying}
S.~Kim, E.~J. Whitehead, and Y.~Zhang, ``Classifying software changes: Clean or
  buggy?'' Software Engineering, IEEE Transactions on, vol.~34, no.~2, 2008,
  pp. 181--196.

\bibitem{ren2004chianti}
X.~Ren, F.~Shah, F.~Tip, B.~G. Ryder, and O.~Chesley, ``Chianti: a tool for
  change impact analysis of java programs,'' in ACM Sigplan Notices, vol.~39,
  no.~10.\hskip 1em plus 0.5em minus 0.4em\relax ACM, 2004, pp. 432--448.

\bibitem{orso2003leveraging}
A.~Orso, T.~Apiwattanapong, and M.~J. Harrold, ``Leveraging field data for
  impact analysis and regression testing,'' in ACM SIGSOFT Software Engineering
  Notes, vol.~28, no.~5.\hskip 1em plus 0.5em minus 0.4em\relax ACM, 2003, pp.
  128--137.

\bibitem{zhang2011localizing}
L.~Zhang, M.~Kim, and S.~Khurshid, ``Localizing failure-inducing program edits
  based on spectrum information,'' in Software Maintenance (ICSM), 2011 27th
  IEEE International Conference on.\hskip 1em plus 0.5em minus 0.4em\relax
  IEEE, 2011, pp. 23--32.

\bibitem{schroter2010stack}
A.~Schroter, N.~Bettenburg, and R.~Premraj, ``Do stack traces help developers
  fix bugs?'' in Mining Software Repositories (MSR), 2010 7th IEEE Working
  Conference on.\hskip 1em plus 0.5em minus 0.4em\relax IEEE, 2010, pp.
  118--121.

\bibitem{nir2008locating}
D.~Nir, S.~Tyszberowicz, and A.~Yehudai, ``Locating regression bugs,'' in
  Hardware and Software: Verification and Testing.\hskip 1em plus 0.5em minus
  0.4em\relax Springer, 2008, pp. 218--234.

\bibitem{miller1995wordnet}
G.~A. Miller, ``Wordnet: a lexical database for english,'' Communications of
  the ACM, vol.~38, no.~11, 1995, pp. 39--41.

\bibitem{stackoverflow}
``Stack overflow,'' \url{http://stackoverflow.com/}.

\bibitem{wala}
``Ibm. the t.j. watson libraries for analysis (wala),''
  \url{http://wala.sourceforge.net/}.

\bibitem{appleton1998streamed}
B.~Appleton, S.~Berczuk, R.~Cabrera, and R.~Orenstein, ``Streamed lines:
  Branching patterns for parallel software development,'' in Proceedings of
  PloP, vol.~98.\hskip 1em plus 0.5em minus 0.4em\relax Citeseer, 1998.

\bibitem{eclipse}
``Eclipse ide,'' \url{https://www.eclipse.org}.

\bibitem{emma}
``{EMMA}: a free java code coverage tool,'' \url{http://emma.sourceforge.net}.

\bibitem{eclemma}
``{EclEmma}: Java code coverage for eclipse,'' \url{http://www.eclemma.org}.

\bibitem{usui2005bugdel}
Y.~Usui and S.~Chiba, ``Bugdel: An aspect-oriented debugging system,'' in
  Software Engineering Conference, 2005. APSEC'05. 12th Asia-Pacific.\hskip 1em
  plus 0.5em minus 0.4em\relax IEEE, 2005, pp. 6--pp.

\bibitem{chiba1998javassist}
S.~Chiba, ``Javassist—a reflection-based programming wizard for java,'' 1998.

\bibitem{angularjs}
``{AngularJS}: An open-source web application framework,''
  \url{https://angularjs.org/}.

\bibitem{xie2005checking}
T.~Xie and D.~Notkin, ``Checking inside the black box: Regression testing by
  comparing value spectra,'' Software Engineering, IEEE Transactions on,
  vol.~31, no.~10, 2005, pp. 869--883.

\bibitem{poshyvanyk2007feature}
D.~Poshyvanyk, Y.-G. Gu{\'e}h{\'e}neuc, A.~Marcus, G.~Antoniol, and V.~Rajlich,
  ``Feature location using probabilistic ranking of methods based on execution
  scenarios and information retrieval,'' Software Engineering, IEEE
  Transactions on, vol.~33, no.~6, 2007, pp. 420--432.

\bibitem{qi2012darwin}
D.~Qi, A.~Roychoudhury, Z.~Liang, and K.~Vaswani, ``Darwin: An approach to
  debugging evolving programs,'' ACM Transactions on Software Engineering and
  Methodology (TOSEM), vol.~21, no.~3, 2012, p.~19.

\bibitem{jagannathantech}
K.~H. P. E.~S. Jagannathan, ``Tech report dynt-200811-1 rprism: Efficient
  regression analysis using view-based trace differencing,'' 2008.

\bibitem{wilde1995software}
N.~Wilde and M.~C. Scully, ``Software reconnaissance: mapping program features
  to code,'' Journal of Software Maintenance: Research and Practice, vol.~7,
  no.~1, 1995, pp. 49--62.

\bibitem{zeller1999yesterday}
A.~Zeller, ``Yesterday, my program worked. today, it does not. why?'' in
  Software Engineering—ESEC/FSE’99.\hskip 1em plus 0.5em minus 0.4em\relax
  Springer, 1999, pp. 253--267.

\bibitem{jss}
\BIBentryALTinterwordspacing
K.~Yu, M.~Lin, J.~Chen, and X.~Zhang, ``Towards automated debugging in software
  evolution: Evaluating delta debugging on real regression bugs from the
  developers’ perspectives,'' Journal of Systems and Software, vol.~85, Oct
  2012, pp. 2305--2317. [Online]. Available:
  \url{http://ast.nlsde.buaa.edu.cn/~kaiyu/pub/JSS.pdf}
\BIBentrySTDinterwordspacing

\bibitem{pastore2012dynamic}
F.~Pastore, L.~Mariani, A.~Goffi, M.~Oriol, and M.~Wahler, ``Dynamic analysis
  of upgrades in c/c++ software,'' in Software Reliability Engineering (ISSRE),
  2012 IEEE 23rd International Symposium on.\hskip 1em plus 0.5em minus
  0.4em\relax IEEE, 2012, pp. 91--100.

\bibitem{ernst2007daikon}
M.~D. Ernst, J.~H. Perkins, P.~J. Guo, S.~McCamant, C.~Pacheco, M.~S. Tschantz,
  and C.~Xiao, ``The daikon system for dynamic detection of likely
  invariants,'' Science of Computer Programming, vol.~69, no.~1, 2007, pp.
  35--45.

\bibitem{zhang2013injecting}
L.~Zhang, L.~Zhang, and S.~Khurshid, ``Injecting mechanical faults to localize
  developer faults for evolving software,'' in Proceedings of the 2013 ACM
  SIGPLAN international conference on Object oriented programming systems
  languages \& applications.\hskip 1em plus 0.5em minus 0.4em\relax ACM, 2013,
  pp. 765--784.

\bibitem{alvesfault}
E.~Alves, M.~Gligoric, V.~Jagannath, and M.~d’Amorim, ``Fault-localization
  using dynamic slicing and change impact analysis.''

\bibitem{iregression_ase12}
\BIBentryALTinterwordspacing
K.~Yu, M.~Lin, J.~Chen, and X.~Zhang, ``Practical isolation of failure-inducing
  changes for debugging regression faults,'' in ASE 2012: Proceedings of the
  27th International Conference on Automated Software Engineering, 2012, pp.
  45--54. [Online]. Available:
  \url{http://ast.nlsde.buaa.edu.cn/~kaiyu/pub/ASE12.pdf}
\BIBentrySTDinterwordspacing

\bibitem{weissgerber2006identifying}
P.~Weissgerber and S.~Diehl, ``Identifying refactorings from source-code
  changes,'' in Automated Software Engineering, 2006. ASE'06. 21st IEEE/ACM
  International Conference on.\hskip 1em plus 0.5em minus 0.4em\relax IEEE,
  2006, pp. 231--240.

\end{thebibliography}

\end{document}